\documentclass[10pt,letterpaper]{article}
\usepackage{opex3}
\usepackage{amsmath}

\begin{document}

\title{Novel perspectives for the application of total internal reflection microscopy}

\author{Giovanni Volpe,$^{1,2,*}$ Thomas Brettschneider,$^{2}$ Laurent Helden,$^{2}$ and Clemens Bechinger$^{1,2}$}

\address{$^{1}$Max-Planck-Institut f\"{u}r Metallforschung, Heisenbergstra{\ss}e 3, 70569 Stuttgart, Germany}
\address{$^{2}$2. Physikalisches Institut, Universit\"{a}t Stuttgart, Pfaffenwaldring 57, 70550 Stuttgart, Germany}

\email{*g.volpe@physik.uni-stuttgart.de}
\homepage{http://www.pi2.uni-stuttgart.de/contact/}

\begin{abstract}
Total Internal Reflection Microscopy (TIRM) is a sensitive
non-invasive technique to measure the interaction potentials between
a colloidal particle and a wall with
femtonewton resolution. The equilibrium distribution of the
particle-wall separation distance $z$ is sampled monitoring the
intensity $I$ scattered by the Brownian particle under evanescent
illumination. Central to the data analysis is the knowledge of the
relation between $I$ and the corresponding $z$,
which typically must be known {\it a priori}. This poses considerable
constraints to the experimental conditions where TIRM can be applied (short penetration depth of the evanescent wave, transparent surfaces).
Here, we introduce a method
to experimentally determine $I(z)$ by relying only on the
distance-dependent particle-wall hydrodynamic interactions. We
demonstrate that this method largely extends the range of conditions
accessible with TIRM, and even allows measurements on highly
reflecting gold surfaces where multiple reflections lead to a complex $I(z)$.
\end{abstract}




\section{Introduction}

Total Internal Reflection Microscopy (TIRM)
\cite{Walz1997,Prieve1999} is a fairly new technique to optically
measure the interactions between a single colloidal particle and a
surface using evanescent light scattering. The distribution of
the separation distances sampled by the particle's Brownian motion is
used to obtain the potential energy profile $U(z)$ of the
particle-surface interactions with sub-$k_BT$ resolution, where $k_BT$ is the thermal energy. Amongst
various techniques available to probe the mechanical properties of
microsystems, the strength of TIRM lies in its sensitivity to very
weak interactions. Atomic Force Microscopy (AFM) \cite{Binnig1986,duc91}
requires a macroscopic cantilever as a probe and is typically
limited to forces down to several piconewton ($10^{-12}\, N$); the
sensitivity of Photonic Force Microscopy (PFM)
\cite{Ghislain1993,Berg-sorensen2004,Volpe2007A} can even reach a
few femtonewtons ($10^{-15}\, N$), but this method is usually
applied to bulk measurements far from any surface. TIRM, instead,
can measure forces with femtonewton resolution acting on a particle
near a surface. Over the last years, TIRM has been successfully
applied to study electrostatic \cite{bik90,Grunberg2001}, van der
Waals \cite{bev99}, depletion
\cite{Pie02b,Bevan2002,Helden2003,Kleshchanok2006}, magnetic
\cite{Blickle2005}, and, rather recently, critical Casimir
\cite{Hertlein2008CC} forces.

\begin{figure}[htbp]
\centering\includegraphics[width=8cm]{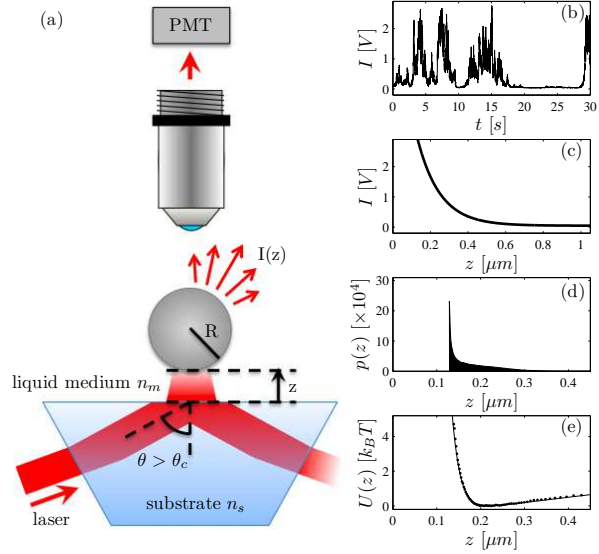} \caption{Total
Internal Reflection Microscopy (TIRM). (a) Schematic of a typical
TIRM setup: a Brownian particle moves in the evanescent
electromagnetic field generated by total internal reflection of a
laser beam; its scattering is collected by an objective lens; and
the scattering intensity is recorded using a photomultiplier (PMT).
(b) Typical experimental scattering intensity time-series
(polystyrene particle in water, $R = 1.45\, \mu m$). (c) Exponential
intensity-distance relation ($\beta = 120\, nm$). (d) Particle
position distribution (acquisition time $1200\,s$, sampling rate
$500\, Hz$). (e) Experimental (dots) and theoretical (line)
potential obtained from the position distribution using the Boltzmann
factor.}\label{Fig1}
\end{figure}

A schematic sketch of a typical TIRM setup is presented in Fig.
\ref{Fig1}(a). To track the Brownian trajectory of a spherical
colloidal particle diffusing near a wall, an evanescent field is
created at the substrate-liquid interface. The scattered light is
collected with a microscope objective and recorded with a
photomultiplier connected to a data acquisition system. Fig.
\ref{Fig1}(b) shows a typical example of an experimentally measured
intensity time-series $I_t$ of a polystyrene particle with radius
$R=1.45\, \mu m$ in water.

Due to the evanescent illumination, the intensity of the light
scattered by the particle is quite sensitive to the
particle-wall distance. If the corresponding intensity-distance
relation $I(z)$ is known (and monotonic), the vertical component of
the particles trajectory $z_t$ can be deduced from $I_t$. To obtain
$I(z)$, it is in principle required to solve a rather complex Mie
scattering problem, i.e. the scattering of a micron-sized colloidal
particle under evanescent illumination close to a surface
\cite{Chew1979,liu95}, where multiple reflections between the particle
and the substrate and Mie resonances must be accounted for
\cite{wan99,liu00,Helden2006,rie07}. When such effects can be
neglected, the scattering intensity is proportional to the
evanescent field intensity and, since the latter decays
exponentially, TIRM data are typically analyzed using a purely
exponential $I(z) = I_0 e^{-z/\beta}$
\cite{Walz1997,Prieve1999,Chew1979,liu95,pri93} (Fig.
\ref{Fig1}(c)), where $\beta = \lambda/4\pi\sqrt{n_s^2\sin^2\theta -
n_m^2}$ is the evanescent field penetration depth,
$\lambda$ the incident light wavelength, $n_s$ the substrate
refractive index, $n_m$ the liquid medium refractive index, and
$\theta$ the incidence angle, which must be larger than the critical
angle $\theta_c = \arcsin(n_m/n_s)$. 
$I_0$ is the scattering intensity at the wall,
which can be determined e.g. using a hydrodynamic method proposed \cite{Bevan2000}.

From the obtained $z_t$ the particle-wall interaction potential,
i.e. $U(z)$, is easily derived by applying the Boltzmann factor $U(z)
= -k_B T \ln{p(z)}$ to the calculated position distribution $p(z)$
(Fig. \ref{Fig1}(d,e)). For an electrically charged dielectric
particle suspended in a solvent, the interaction
potential typically corresponds to $U(z) = B\exp{(-\kappa
z)}+[\frac{4}{3}\pi R^3 (\rho_{p}-\rho_{m})g - F_s] z$. The first
term is due to double layer forces with $\kappa^{-1}$ the Debye
length and $B$ a prefactor depending on the surface charge densities
of the particle and the wall
\cite{Walz1997,Prieve1999,Grunberg2001}. The second term describes
the effective gravitational contributions with $\rho_{p}$ and $\rho_{m}$
the particle and solvent density and $g$ the gravitational
acceleration constant; $F_s$ takes into account additional optical
forces, which may result from a vertically incident laser beam often
employed as a two-dimensional optical trap to reduce the lateral motion of the particle \cite{wal92}.
Depending on the experimental conditions, additional interactions,
such as depletion or van der Waals forces, may arise.

Despite the broad range of phenomena that have successfully been
addressed with TIRM, most studies have been carried out with small
penetration depths (at most $\beta \approx 100\, nm$), and have therefore been
limited to rather small
particle-substrate distances $z$. In addition, no TIRM studies on
highly reflecting walls, e.g. gold surfaces, have been reported,
although such surfaces are interesting since they can support
surface plasmons enhancing the evanescent field \cite{Marti1993} and
the optical near-field radiation forces
\cite{Volpe2006A,Righini2007,Righini2008}. Furthermore, gold
coatings can be easily functionalized \cite{ulm96}, which would allow to apply
TIRM to e.g. biological systems. The reason for these
limitations is the aforementioned problem to obtain a reliable
$I(z)$ relationship under these conditions. For example, it has been
demonstrated that large penetration depths (e.g. above $\approx
200\, nm$ in Ref. \cite{Hertlein2008TIRM}) increase the multiple
optical reflections between the particle and the wall, which in turn
leads to a non-exponential $I(z)$
\cite{Helden2006,Hertlein2008TIRM}. Experiments combining TIRM and
AFM found deviations from simple exponential behaviour very close to the wall 
even for shorter penetration depths \cite{McKee2005}. In principle, such effects can be included into
elaborate scattering models, however, this requires precise
knowledge of the system properties and, in particular, of the
refractive indices of particle, wall, and liquid medium
\cite{Hertlein2008TIRM}. Since the latter are prone to significant
uncertainties (in particular for the colloidal particles), the application of TIRM under such conditions remains inaccurate.

Here, we introduce a method to experimentally determine $I(z)$ by
making solely use of the experimentally acquired $I_t$ and of the
distance-dependent hydrodynamic
interactions between the particle and the wall. In particular, no
knowledge about the shape of the potential $U(z)$ is required. We
demonstrate the capability of this method by experiments and
simulations, and we also apply it to experimental conditions with long
penetration depths ($\beta= 720\,nm$) and even with highly reflective
gold surfaces.

\section{Theory}

\subsection{Diffusion coefficient and skewness of Brownian motion near a wall}

\begin{figure}[htbp]
\centering\includegraphics[width=7.5cm]{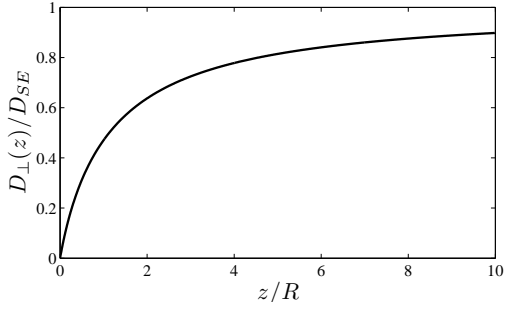} \caption{Vertical diffusion
coefficient  $D_{\bot}(z)$ near a wall (Eq. (\ref{Brennerformula})).}\label{Fig2}
\end{figure}

Colloidal particles immersed in a solvent undergo Brownian motion
due to collisions with solvent molecules. This erratic motion leads
to particle diffusion with the Stokes-Einstein diffusion coefficient
$D_{SE} = k_BT/6\pi\eta R$, where $\eta$ is the shear viscosity of the
liquid. It is well known that this bulk diffusion coefficient
decreases close to a wall due to hydrodynamic interactions. From the
solution of the creeping flow equations for a spherical particle in
motion near a wall assuming nonslip boundary conditions and
negligible inertial effects, one obtains for the diffusion
coefficient in the vertical direction \cite{Brenner1961},
\begin{equation}\label{Brennerformula}
D_{\bot}(z) = \frac{D_{SE}}{l(z)} \mbox{,}
\end{equation}
where $l(z) = \frac{4}{3}\sinh{\left( \alpha(z) \right)}
\sum_{n=1}^{\infty} \frac{n(n+1)}{(2n-1)(2n+3)} \left[ \frac{
2\sinh{\left( (2n+1)\alpha(z) \right)} + (2n+1)\sinh{\left(
2\alpha(z) \right)} }{ 4\sinh^2{\left( (n+0.5)\alpha(z) \right)} -
(2n+1)^2\sinh^2{\left( \alpha(z) \right)}} -1 \right]$ and
$\alpha(z) = \cosh^{-1}\left( 1+\frac{z}{R}\right)$. As shown in
Fig. \ref{Fig2}, $D_{\bot}$ first increases with $z$
approaching the corresponding bulk value at a distance of several
particle radii away from the wall.

Experimentally, the diffusion coefficient can be obtained from the
mean square displacement (MSD) calculated from a particle
trajectory. For the $z$-component this reads ${\langle (z_{t+\Delta
t} - z_t)^2\rangle} = 2 D_{SE} \Delta t$, where $\langle ...
\rangle$ indicates average over time $t$. To account for a
$z$-dependent diffusion coefficient close to a wall, one has to
calculate the conditional MSD given that the particle is at time $t$
at position $z$, i.e. $\langle (z_{t+\Delta t} - z_t)^2 \mid
z_t=z \rangle = 2 D_{\bot}(z) \Delta t$ where the equality is valid for $\Delta t \rightarrow 0$; in such limit, 
this expression is only
determined by the particle diffusion even if the particle is exposed
to an external potential $U(z)$. From this follows that
$D_{\bot}(z)$ can be directly obtained from the particle's
trajectory
\begin{equation}\label{Z_Brenner}
D_{\bot}(z) = \lim_{\Delta t \rightarrow 0} \frac{1}{2\Delta t}
\left\langle (z_{t+\Delta t} - z_t)^2 \mid z_t=z \right\rangle
\mbox{.}
\end{equation}
Eq. (\ref{Z_Brenner}) was employed already by several groups
\cite{Oetama2005,Carbajal-Tinoco2007} to validate Eq. (\ref{Brennerformula}).

The distribution of particle displacements $h(z; z_0,\Delta t)$ around a
given distance $z_0$ converges to a gaussian for $\Delta t
\rightarrow 0$ and therefore its {\it skewness}
-- i.e. the normalized third central moment --
converges to zero. Accordingly,
\begin{equation}\label{Z_Skewness}
S (z) \equiv \lim_{\Delta t \rightarrow 0} \frac{1}{\Delta t^2}
\left\langle \left(
\frac{z_{t+\Delta t} - z_t - M (z, \Delta t)}{ \sqrt{2D_{\bot}(z)} }
\right)^3 \mid z_t=z \right\rangle = 0 \mbox{,}
\end{equation}
where $M (z, \Delta t) = \langle z_{t+\Delta t} - z_t \mid z_t=z \rangle = \underset{\hat{z}}{\arg \max} \; h(\hat{z}; z, \Delta t)$,
where $\underset{\hat{z}}{\arg \max}$ indicates the argument that maximize the given function.

\subsection{Mean square displacement and skewness of the scattering intensity}

\begin{figure}[htbp]
\centering\includegraphics[width=7.5cm]{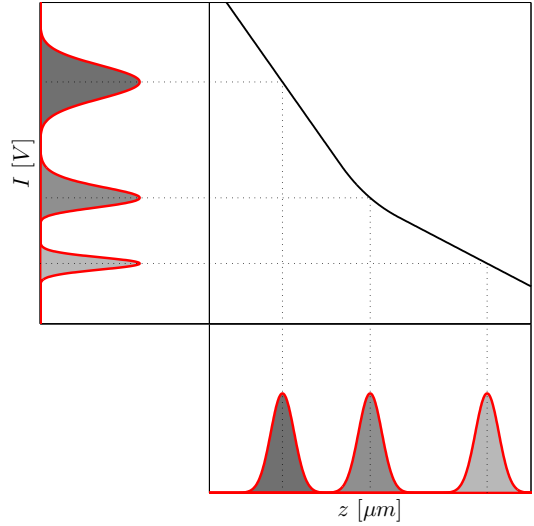} \caption{
Relation between position distributions and intensity distributions. 
Brownian diffusion of a particle around a point is symmetric,
leading for small $\Delta t$ to a gaussian distribution $h(z;z_0,\Delta t)$ (bottom). According to $I(z)$ this leads
to the scattered intensity histograms $h(I;I_0,\Delta t)$
(left): in the linear region of $I(z)$, $h(I;I_0,\Delta t)$
is also gaussian with the width depending on $I'$ (Eq.
(\ref{I_MSD})); in the nonlinear region of $I(z)$ (central histogram),
$h(I;I_0,\Delta t)$ deviates from a gaussian and has a
finite skewness depending on $I''$ (Eq. (\ref{I_Skewness})).}\label{Fig3}
\end{figure}

In a TIRM experiment, $h(z; z_0, \Delta t)$ is translated
into a corresponding intensity distribution $h(I; I_0, \Delta t)$ around
intensity $I_0 = I(z_0)$, whose shape strongly depends on $I(z)$. In Fig. \ref{Fig3} we demonstrate how a particle displacement distribution
$h(z; z_0, \Delta t)$, which is gaussian for small $\Delta t$, translates into the corresponding scattered intensity distribution $h(I; I_0, \Delta t)$ for an arbitrary non-exponential $I(z)$ dependence. In the linear regions of $I(z)$,
the corresponding $h(I; I_0, \Delta t)$ are also gaussian with the half-width
determined by the slope of the $I(z)$ curve. In the non-linear part
of $I(z)$, however, a non-gaussian intensity histogram with finite
skewness is obtained.

In the following we calculate the MSD and the skewness of $h(I; I_0, \Delta t)$
for an arbitrary $I(z)$, which we assume to be a continuous function with well defined first and second derivates $I'$ and $I''$.
In the vicinity of $z_0$, $I(z)$ can be therefore expanded in a Taylor series
$I(z) = I(z_0) + I'(z_0)(z-z_0) + \frac{1}{2} I''(z_0)(z-z_0)^2$ for $z \rightarrow z_0$, where $I' = \frac{dI}{dz}$ and $I'' =
\frac{d^2I}{dz^2}$. The MSD of $h(I; I_0, \Delta t)$ is
\begin{equation}\label{I_MSD}
\mathrm{MSD}(I) \equiv \lim_{\Delta t \rightarrow 0} \frac{1}{\Delta
t} \left\langle (I_{t+\Delta t}-I_t)^2 \mid I_t = I \right\rangle =
I'^2 (z) \cdot 2 D_{\bot} (z) \mbox{,}
\end{equation}
where $\left\langle (I_{t+\Delta t}-I_t)^2 \mid I_t = I
\right\rangle = I'^2 \left\langle (z_{t+\Delta t}-z_t)^2 \mid I_t =
I \right\rangle$ for $\Delta t \rightarrow 0$ and Eq. (\ref{Z_Brenner}) has been
used. The skewness of $h(I; I_0, \Delta t)$ is
\begin{equation}\label{I_Skewness}
\mathrm{S}(I) \equiv \lim_{\Delta t \rightarrow 0} \frac{1}{\Delta
t^2} \left\langle \left( \frac{I_{t+\Delta t} - I_t - M(I,\Delta t)}{ \sqrt{\mathrm{MSD}(I)} }
\right)^3 \mid I_t = I \right\rangle = \frac{9}{2} \frac{I''(z)}{|I'(z)|} \cdot
\sqrt{2 D_{\bot}(z)} \mbox{,}
\end{equation}
with $M(I,\Delta t) = \underset{\hat{I}}{\arg \max} \; h(\hat{I}; I, \Delta t)$ 
and 
$\left\langle (I_{t+\Delta t} - I_t - M(I,\Delta t))^3 \mid I_t = I \right\rangle 
= I'^3 \left\langle (z_{t+\Delta t} - z_t - M(z,\Delta t))^3 \mid z_t = z \right\rangle 
+ \frac{3}{2} I'^2 I'' \left\langle (z_{t+\Delta t} - z_t - M(z,\Delta t) )^4 \mid z_t = z \right\rangle$ for $\Delta
t \rightarrow 0$ where the first term is null because of Eq.
(\ref{Z_Skewness}), and the second term is calculated using the
properties of the momenta of a gaussian distribution $\left\langle
(...)^4 \right\rangle = 3 \left\langle (...)^2 \right\rangle^2$.

\begin{figure}[h]
\centering\includegraphics[width=8cm]{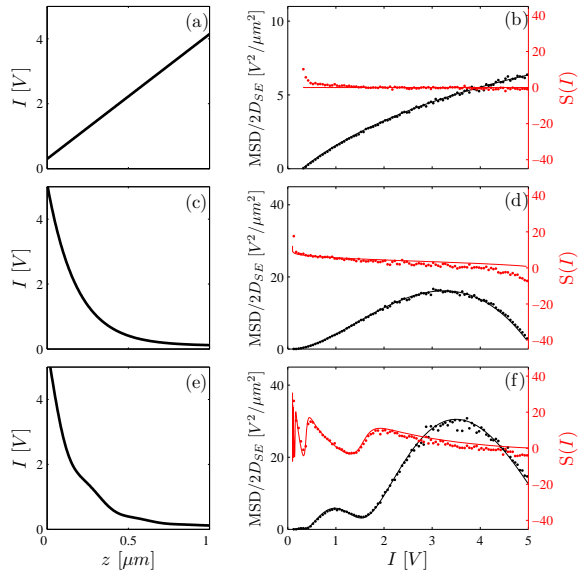} \caption{Various
intensity-distance relations and their effect on 
$\mathrm{MSD}(I)$ (black) and skewness $\mathrm{S}(I)$
(red). Both theoretical values (solid lines) and the results from
the analysis of numerically simulated data (dots) using Eq.
(\ref{I_MSD}) and Eq. (\ref{I_Skewness}) are presented ($R = 1.45\,
\mu m$, $\rho_p = 1.053\, g/cm^3$, samples $10^6$, frequency
$100\, Hz$). (a)-(b) Linear $I(z)$. (c)-(d) Exponential $I(z)$.
(e)-(f) Exponential $I(z)$ modulated by a sinusoidal
function.}\label{Fig4}
\end{figure}

In Fig. \ref{Fig4} we applied Eqs. (\ref{I_MSD}) and (\ref{I_Skewness})
to the intensity time-series corresponding to a particle trajectory simulated 
using a Langevin difference equation assuming various $I(z)$ (Fig. \ref{Fig4}(a,c,e)).
The solid lines in Fig. \ref{Fig4}(b,d,f) show the theoretical $\mathrm{MSD}(I)$ (black) and $\mathrm{S}(I)$ (red) 
and the dots the ones obtained from the simulations.
When $I(z)$ is linear (Fig.
\ref{Fig4}(a)), $\mathrm{MSD}(I)$ is proportional to Eq.
(\ref{Brennerformula}) and $\mathrm{S}(I)$ vanishes (Fig. \ref{Fig4}(b)).
When $I(z)$ is exponential (Fig. \ref{Fig4}(c)) or a sinusoidally
modulated exponential (Fig. \ref{Fig4}(e), $\mathrm{MSD}(I)$ is not proportional to Eq.
(\ref{Brennerformula}) and large values of the skewness occur as shown in (Figs. \ref{Fig4}(d,f)). 
Small deviations between the theoretical curves and the numerical data can be observed
for intensities where the particle drift becomes large in comparison to the time-step ($\Delta t=2\, ms$);
in our case this corresponds to a slope of the potential of about $1\,pN/\mu m$, which is close
to the upper force limit of typical TIRM measurements. If necessary
such deviations can be reduced employing shorter time-steps.

\subsection{Obtaining $I(z)$ from $I_t$}

The correct $I(z)$ satisfies the conditions
\begin{equation}\label{conditions}
\left\{
\begin{array}{ccc}
\mathrm{MSD}(I(z)) & = & I'^2(z) \cdot 2 D_{\bot}(z) \\
\mathrm{S}(I(z)) & = &\frac{9}{2} \frac{I''(z)}{|I'(z)|} \cdot \sqrt{2 D_{\bot}(z)}
\end{array}
\right.
\mbox{,}
\end{equation}
where $\mathrm{MSD}(I)$ and $\mathrm{S}(I)$ are calculated from an
experimental $I_t$. Thus, the problem of determining $I(z)$ can be regarded
as a functional optimization problem, where Eqs. (\ref{conditions}) have to be fulfilled.

\section{Analysis workflow}

Here, we present a concrete analysis workflow to obtain $z_t$ from the experimental $I_t$ by
finding the $I(z)$ that satisfies Eqs. (\ref{conditions}). To do so, we
will construct a series of approximations $I^{(i)}(z)$ indexed by
$i$ converging to $I(z)$.

(1) As first guess, take $I^{(0)} (z) = I_0 \exp{(-z/\beta)}+b_s$, where $\beta$,
$I_0$ and $b_s$ are parameters chosen to optimize Eqs. (\ref{conditions}).
Often some initial estimates are available from the experimental conditions:
$\beta$ can be taken as the evanescent field penetration depth, $I_0$ as
the scattering intensity at the wall, and $b_s$ as the background
scattering in the absence of the Brownian particle.
While $\beta$ is typically well known, $I_0$ and $b_s$ are prone to large
experimental systematic errors and uncertainties.

(2) Take $I^{(1)} (z) = I^{(0)}(z) [1-G(I^{(0)}(z), \mu^{(1)},
\sigma^{(1)}, A^{(1)}) ]$, where  $G(x, \mu, \sigma, A) = A
\exp(-(x-\mu)^2/\sigma^2)$ is a gaussian, and the parameters $\mu^{(1)}$,
$\sigma^{(1)}$, and $A^{(1)}$ optimize Eqs.  (\ref{conditions}).
Gaussian functions were chosen because they have smooth derivatives and quickly tend to zero at infinite. 
Notice that the choice of a Gaussian is unessential for the working of the algorithm.

(3) Reiterate step (2), substituting $I^{(0)}$ with $I^{(i)}$ and
$I^{(1)}$ with $I^{(i+1)}$, until Eqs. (\ref{conditions}) are satisfied within the required precision.

(4) Invert $I^{(i+1)}(z)$, i.e. numerically construct $z^{(i+1)}(I)$.

(5) Take $z_t= z^{(i+1)}(I_t)$.

\section{Experimental case studies}

\subsection{Validation of the technique}

\begin{figure}[h]
\centering\includegraphics[width=8cm]{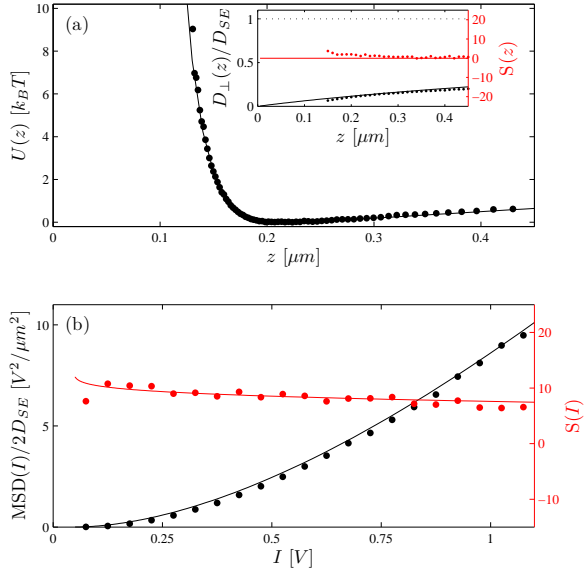} \caption{ TIRM with
exponential intensity-distance relation. (a) The experimental (dots)
and theoretical (solid line) potential. Inset: the diffusion
coefficient on the position data (black dots) fits well Eq.
(\ref{Brennerformula}) (black solid line), while the absolute value
of the Brownian motion skewness $\mathrm{S}(z)$ is small (red dots). (b)
Experimental $\mathrm{MSD}(I)$ (black dots) and skewness $\mathrm{S}(I)$ (red
dots) for a scattering intensity time-series (polystyrene particle,
$R = 1.45\, \mu m$, $\rho_p = 1.053\, g/cm^3$, $n_p=1.59$ suspended in water $n_m = 1.33$ with $300\, \mu M$
NaCl background electrolyte, $\kappa^{-1} = 17\, nm$,
near a glass surface $n_s=1.52$, acquisition time $1800\, s$,
sampling frequency $500 \, Hz$) calculated using Eq. (\ref{I_MSD})
and Eq. (\ref{I_Skewness}). Given the short penetration depth
($\beta = 120\, nm$), the theoretical $\mathrm{MSD}(I)$ (black solid
line) and $\mathrm{S}(I)$ (red solid line) for an exponential $I(z)$
fit the experimental ones and the conditions in Eqs.
(\ref{conditions}) are fulfilled. }\label{Fig5}
\end{figure}

We test our approach on experimental data (polystyrene particle with
$R = 1.45\,\mu m$ near a glass-water interface kept in place by a
vertically incident laser beam \cite{wal92}) for which the
exponential $I(z)$ is justified ($\beta = 120\, nm$, $\lambda = 658
\, nm$) \cite{Hertlein2008TIRM}. As illustrated in Fig.
\ref{Fig5}(a), there is indeed agreement between the measured (dots)
and theoretical potential (solid line). In the inset, the measured
diffusion coefficient (black dots) agrees with Eq. (\ref{Z_Brenner})
(black solid line) and the skewness (red dots) is negligible (small
deviations in the region where the potential is steepest are due to
the finite time-step). The criteria for $I(z)$ in Eqs.
(\ref{conditions}) are already fulfilled after $I_0$ and $b_s$ have
been optimized in the first step of the analysis workflow in the
previous section. As shown in Fig. \ref{Fig5}(b), the experimental
$\mathrm{MSD}(I)$ (black dots) and skewness $\mathrm{S}(I)$ (red dots) fit Eqs. (\ref{I_MSD}) and
(\ref{I_Skewness}) (solid lines).

\begin{figure}[t]
\centering\includegraphics[width=8cm]{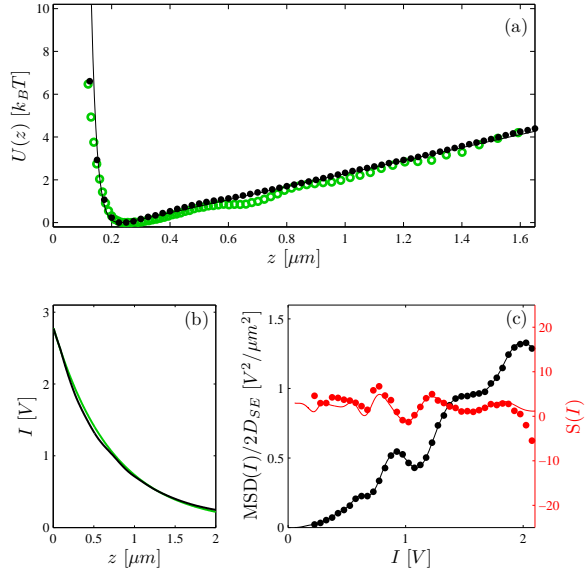} \caption{TIRM with
large penetration depth. (a) The experimental potential (black dots)
obtained using the fitted intensity-distance relation and the
theoretical one (black solid line). The green dots represent the
faulty potential obtained using the exponential $I(z)$. (b) The
fitted $I(z)$ (black line) and the exponential one (green line)
corresponding to the penetration depth $\beta =
720\, nm$. (c) Experimental intensity $\mathrm{MSD}(I)$ (black dots) and
skewness $\mathrm{S}(I)$ (red dots) for a scattering intensity time-series
(same particle and acquisition parameters as in Fig. \ref{Fig5})
calculated using Eq. (\ref{I_MSD}) and Eq. (\ref{I_Skewness}). Due
to the large penetration depth, the $I(z)$ diverges from an
exponential; the theoretical $\mathrm{MSD}(I)$ (black solid line)
and $\mathrm{S}(I)$ (red solid line) correspond to the fitted
$I(z)$.}\label{Fig6}
\end{figure}

\begin{figure}[t]
\centering\includegraphics[width=8cm]{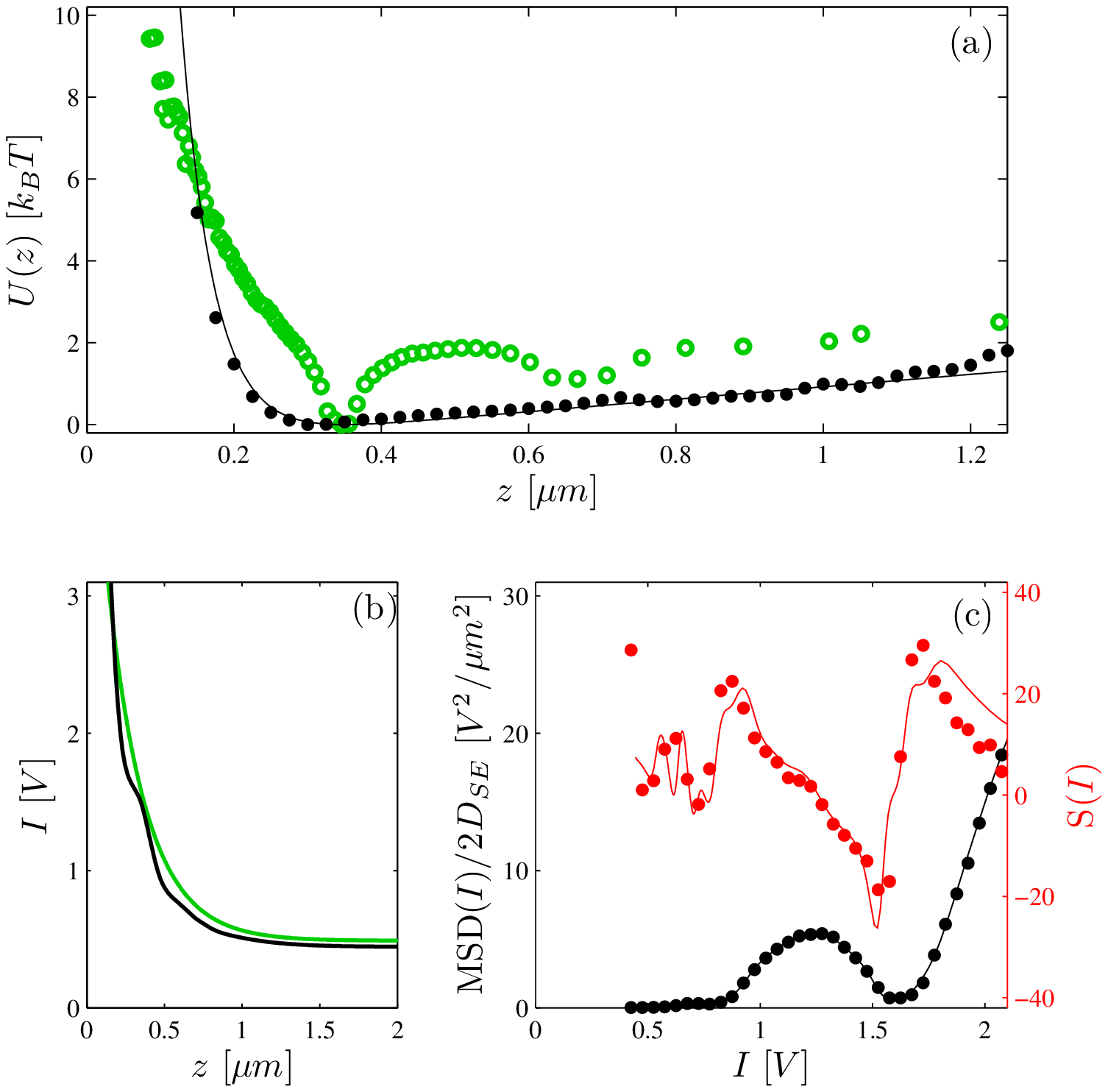} \caption{TIRM in
front of a reflective surface. (a) The experimental potential (black
dots) obtained using the fitted intensity-distance relation and
theoretical one (solid black line). The green dots represent the
faulty potential obtained using the exponential $I(z)$ with $\beta =
244\, nm$. (b) The fitted $I(z)$ (black line) and the exponential
one (green line) corresponding to the evanescent field penetration
depth $\beta = 244\, nm$. (c) Experimental intensity
$\mathrm{MSD}(I)$ (black dots) and skewness $\mathrm{S}(I)$ (red dots) for a
scattering intensity time-series (same particle and acquisition
parameters as in Fig. \ref{Fig5}, except for background electrolyte
$50\, \mu M$, $\kappa^{-1} = 42\, nm$)  calculated
using Eq. (\ref{I_MSD}) and Eq. (\ref{I_Skewness}). Due to the
presence of a $20\, nm$-thick gold layer on the surface, the $I(z)$
deviates from an exponential; the theoretical $\mathrm{MSD}(I)$
(black solid line) and $\mathrm{S}(I)$ (red solid line) correspond
to the fitted $I(z)$. }\label{Fig7}
\end{figure}

\subsection{TIRM with large penetration depth}

We now apply our technique under conditions where an exponential
$I(z)$ is not valid, i.e. for large penetration depth as mentioned above. 
Fig. \ref{Fig6} shows the potential obtained for the same particle as in Fig. \ref{Fig5} but for a 
penetration depth ($\beta = 720\, nm$). Note, that compared to \ref{Fig5} the potential extends over a much larger distance range since the particle's motion can be tracked from hundreds of
nanometers to microns. The green data points show the faulty interaction potential that is obtained when assuming an exponential $I(z)$.
Since the only difference is in the illumination, the same potential as in Fig. \ref{Fig5}
should be retrieved (solid line in Fig. \ref{Fig6}(a)). However,
applying an exponential $I(z)$ (green line in Fig. \ref{Fig6}(b))
wiggles appear in the potential (green dots in Fig. \ref{Fig6}(a)).
Their origin is due to multiple reflections between the particle and the wall as discussed in detail in \cite{Helden2006}.  The
correct $I(z)$ (black line in Fig. \ref{Fig6}(b)) is obtained with
the algorithm proposed in the previous section: after 9 iterations
the conditions in Eqs. (\ref{conditions}) appear reasonably
satisfied, as shown in Fig. \ref{Fig6}(c). With this $I(z)$, we
reconstructed the potential represented by the black dots in Fig.
\ref{Fig6}(a), in good agreement with the one in Fig. \ref{Fig5}. It should be noticed that, even though the deviations of the correct $I(z)$ from
an exponential function are quite small, this is enough to
significantly alter the measurement of the potential. This again demonstrates the importance of obtaining the correct $I(z)$ for the analysis of TIRM experiments.

\subsection{TIRM in front of a reflective surface}

To demonstrate that our method is capable of correcting even more severe optical
distortions, we performed measurements in front of a reflecting
surface ($20\, nm$ gold-layer, reflectivity $\approx 60\%$, $\beta =
244\, nm$). The experimental conditions are similar to the previous
experiments. Only the salt concentration was lowered to avoid
sticking of the particle to the gold surface due to van der Waals
forces, leading to a larger electrostatic particle-surface repulsion, and 
the optical trap was not used.
Using an exponential $I(z)$ (green
line in Fig. \ref{Fig7}(b)), we obtain the potential represented by
the green dots in Fig. \ref{Fig7}(a), which clearly features
unphysical artifacts, e.g. spurious potential minima. After 27
iterations of the data analysis algorithm, the black $I(z)$ in Fig.
\ref{Fig7}(b) is obtained, which reasonably satisfies the criteria
in Eqs. (\ref{conditions}) (Fig. \ref{Fig7}(c)). The reconstructed
potential (black dots in Fig. \ref{Fig7}(a)) fits well to theoretical predictions (solid line); in particular the
unphysical minima disappear. 

\section{Conclusions \& Outlook}

TIRM is a technique which allows one to measure the interaction potentials between a 
colloidal particle and a wall with femtonewton resolution.
So far, its applicability has been limited by the need for an {\it a priori} knowledge of the intensity-distance relation.
$I(z) \propto \exp(-z/\beta)$ can safely be assumed only for short penetration
depths of the evanescent field and transparent surfaces. 
This, however, poses considerable constraints to the experimental conditions and
the range of forces where TIRM can be applied.
Here, we have proposed a technique to determine $I(z)$ that relies only on the hydrodynamic particle-surface interaction (Eq. \ref{Brennerformula}) and, differently from existing data evaluation schemes, makes no assumption on the functional form of $I(z)$ or on the wall-particle potential.
This technique will particularly be beneficial for the extension of TIRM to new domains.
Here, we have demonstrated TIRM with a very large penetration depth,
which allows one to bridge the gap between surface measurements and bulk measurements,
and TIRM in front of a reflecting (gold-coated) surface,
which allows plasmonic and biological applications.

This new technique only assumes the knowledge of the
particle radius, which is usually known within an high accuracy and
can also be measured {\it in situ} \cite{Bevan2000}, and the
monotonicity of $I(z)$. Were $I(z)$ not monotonous, as it might
happen for a metallic particle in front of a reflective surface, the
technique can be adapted to use the information from two
non-monotonous signals, e.g. the scattering from two evanescent
fields with different wavelength \cite{Hertlein2008TIRM}.
We notice that the technique encounters its natural limits when Eq. (\ref{Brennerformula}) does not correctly describe the particle-wall hydrodynamic interactions. This may happen in situations when the stick boundary conditions do not apply
or when the hydrodynamic interactions are otherwise altered, e.g. in a viscoelastic fluid.

Since the conditions in Eqs. (\ref{conditions}) are fulfilled only
by the correct $I(z)$, they permit a self-consistency check on the
data analysis. Even when an exponential $I(z)$ is justified, errors
that arise from the estimation of some parameters (e.g. the
zero-intensity $I_0$ and the background intensity $b_s$) can be
easily avoided by checking the consistency of the analyzed data with
the aforementioned criteria. In principle, the analysis of TIRM data
can be completely automatized, possibly providing the missing link
for a widespread application of TIRM to fields, such as biology,
where automated analysis techniques are highly appreciated.

The proposed technique can also be useful to determine the
intensity-distance relation in all those situations where it is
possible to rely on the knowledge of the system hydrodynamics, while
the scattering is not accurately known. Often explicit formulas are
available for the hydrodynamic interaction of an over-damped
Brownian particle in a simple geometry, while complex numerical
calculations are needed to determine its scattering. As a limiting
case, this technique might also prove useful for the PFM technique
working in bulk, where the diffusivity is constant. Indeed, under
certain experimental conditions -- e.g. using back-scattered light
instead of the more usual forward-scattered light \cite{Volpe2007B}
-- the intensity-distance relation can be non-trivial and it can be
necessary to determine it experimentally.


\begin{thebibliography}{10}
\newcommand{\enquote}[1]{``#1''}
\expandafter\ifx\csname url\endcsname\relax
  \def\url#1{\texttt{#1}}\fi
\expandafter\ifx\csname urlprefix\endcsname\relax\def\urlprefix{URL }\fi
\providecommand{\eprint}[2][]{\url{#2}}

\bibitem{Walz1997}
J.~Walz, \enquote{Measuring particle interactions with total internal
  reflection microscopy,} Curr. Opin. Colloid. Interface Sci. \textbf{1997},
  600--606 (2).

\bibitem{Prieve1999}
D.~C. Prieve, \enquote{Measurement of colloidal forces with TIRM,} Adv.
  Colloid. Interface Sci. \textbf{82}, 93--125 (1999).

\bibitem{Binnig1986}
G.~Binnig, C.~F. Quate, and C.~Gerber, \enquote{Atomic force microscope,} Phys.
  Rev. Lett. \textbf{56}, 930--933 (1986).

\bibitem{duc91}
W.~Ducker, T.~Senden, and R.~Pashley, \enquote{Direct measurement of colloidal
  forces using an atomic force microscope,} Nature \textbf{353}, 239--241
  (1991).

\bibitem{Ghislain1993}
L.~P. Ghislain and W.~W. Webb, \enquote{Scanning-force microscope based on an
  optical trap,} Opt. Lett. \textbf{18}, 1678--1680 (1993).

\bibitem{Berg-sorensen2004}
K.~Berg-S{\o}rensen and H.~Flyvbjerg, \enquote{Power spectrum analysis for
  optical tweezers,} Rev. Sci. Instrum. \textbf{75}, 594--612 (2004).

\bibitem{Volpe2007A}
G.~Volpe, G.~Volpe, and D.~Petrov, \enquote{Brownian motion in a nonhomogeneous
  force field and photonic force microscope,} Phys. Rev. E \textbf{76}, 061,118
  (2007).

\bibitem{bik90}
S.~G. Bike and D.~C. Prieve, \enquote{Measurements of double-layer repulsion
  for slightly overlapping counterion clouds,} Int. J. Multiphase Flow
  \textbf{16}, 727--740 (1990).

\bibitem{Grunberg2001}
H.~H. von Gr\"{u}nberg, L.~Helden, P.~Leiderer, and C.~Bechinger,
  \enquote{Measurement of surface charge densities on Brownian particles using
  total internal reflection microscopy,} J. Chem. Phys. \textbf{114},
  10,094--10,104 (2001).

\bibitem{bev99}
M.~A. Bevan and D.~C. Prieve, \enquote{Direct measurement of retarded van der
  Waals attraction,} Langmuir \textbf{15}, 7925--7936 (1999).

\bibitem{Pie02b}
M.~Piech, P.~Weronski, X.~Wu, and J.~Y. Walz, \enquote{Prediction and
  measurement of the interparticle depletion interaction next to a flat wall,}
  J. Colloid Interface Sci. \textbf{247}, 327--341 (2002).

\bibitem{Bevan2002}
M.~A. Bevan and P.~J. Scales, \enquote{Solvent quality dependent interactions
  and phase behavior of polystyrene particles with physisorbed PEO-PPO-PEO,}
  Langmuir \textbf{18}, 1474--1484 (2002).

\bibitem{Helden2003}
L.~Helden, R.~Roth, G.~H. Koenderink, P.~Leiderer, and C.~Bechinger,
  \enquote{Direct measurement of entropic forces induced by rigid rods,} Phys.
  Rev. Lett. \textbf{90}, 048,301 (2003).

\bibitem{Kleshchanok2006}
D.~Kleshchanok, R.~Tuinier, and P.~R. Lang, \enquote{Depletion interaction
  mediated by a polydisperse polymer studied with total internal reflection
  microscopy,} Langmuir \textbf{22}, 9121--9128 (2006).

\bibitem{Blickle2005}
V.~Blickle, D.~Babic, and C.~Bechinger, \enquote{Evanescent light scattering
  with magnetic colloids,} Appl. Phys. Lett. \textbf{87}, 101,102 (2005).

\bibitem{Hertlein2008CC}
C.~Hertlein, L.~Helden, A.~Gambassi, S.~Dietrich, and C.~Bechinger,
  \enquote{Direct measurement of critical Casimir forces,} Nature \textbf{451},
  172--175 (2008).

\bibitem{Chew1979}
H.~Chew, D.-S. Wang, and M.~Kerker, \enquote{Elastic scattering of evanescent
  electromagnetic waves,} Appl. Opt. \textbf{18}, 2679--2687 (1979).

\bibitem{liu95}
C.~C.~Liu, T.~Kaiser, S.~Lange, and G.~Schweiger, \enquote{Structural
  resonances in a dielectric sphere illuminated by an evanescent wave,} Opt.
  Commun. \textbf{117}, 521--531 (1995).

\bibitem{wan99}
R.~Wannemacher, A.~Pack, and M.~Quinten, \enquote{Resonant absorption and
  scattering in evanescent fields,} Appl. Phys. B \textbf{68}, 225--232 (1999).

\bibitem{liu00}
C.~Liu, T.~Weigel, and G.~Schweiger, \enquote{Structural resonances in a
  dielectric sphere on a dielectric surface illuminated by an evanescent wave,}
  Opt. Commun. \textbf{185}, 249--261 (2000).

\bibitem{Helden2006}
L.~Helden, E.~Eremina, N.~Riefler, C.~Hertlein, C.~Bechinger, Y.~Eremin, and
  T.~Wriedt, \enquote{Single-particle evanescent light scattering simulations
  for total internal reflection microscopy,} Appl. Opt. \textbf{45}, 7299--7308
  (2006).

\bibitem{rie07}
N.~Riefler, E.~Eremina, C.~Hertlein, L.~Helden, Y.~Eremin, T.~Wriedt, and
  C.~Bechinger, \enquote{Comparison of T-matrix method with discrete sources
  method applied for total internal reflection microscopy,} J. Quant. Spectr.
  Rad. Transfer \textbf{106}, 464--474 (2007).

\bibitem{pri93}
D.~C. Prieve and J.~Y. Walz, \enquote{Scattering of an evanescent surface wave
  by a microscopic dielectric sphere,} Appl. Opt. \textbf{32}, 1629--1641
  (1993).

\bibitem{Bevan2000}
M.~A. Bevan and D.~C. Prieve, \enquote{Hindered diffusion of colloidal
  particles very near to a wall: Revisited,} J. Chem. Phys. \textbf{113},
  1228--1236 (2000).

\bibitem{wal92}
J.~Y. Walz and D.~C. Prieve, \enquote{Prediction and measurement of the optical
  trapping forces on a dielectric sphere,} Langmuir \textbf{8}, 3073--3082
  (1992).

\bibitem{Marti1993}
O.~Marti, H.~Bielefeldt, B.~Hecht, S.~Herminghaus, P.~Leiderer, and J.~Mlynek,
  \enquote{Near-field optical measurement of the surface plasmon field,} Opt.
  Commun. \textbf{96}, 225--228 (1993).

\bibitem{Volpe2006A}
G.~Volpe, R.~Quidant, G.~Badenes, and D.~Petrov, \enquote{Surface plasmon
  radiation forces,} Phys Rev. Lett. \textbf{96}, 238,101 (2006).

\bibitem{Righini2007}
M.~Righini, A.~S. Zelenina, C.~Girard, and R.~Quidant, \enquote{Parallel and
  selective trapping in a patterned plasmonic landscape,} Nat. Phys.
  \textbf{3}, 477--480 (2007).

\bibitem{Righini2008}
M.~Righini, G.~Volpe, C.~Girard, D.~Petrov, and R.~Quidant, \enquote{Surface
  plasmon optical tweezers: tunable optical manipulation in the femto-newton
  range,} Phys. Rev. Lett. \textbf{100}, 186,804 (2008).

\bibitem{ulm96}
A.~Ulman, \enquote{Formation and structure of self-assembled monolayers,} Chem.
  Rev. \textbf{96}, 1533--1554 (1996).

\bibitem{Hertlein2008TIRM}
C.~Hertlein, N.~Riefler, E.~Eremina, T.~Wriedt, Y.~Eremin, L.~Helden, and
  C.~Bechinger, \enquote{Experimental verification of an exact evanescent light
  scattering model for TIRM,} Langmuir \textbf{24}, 1--4 (2008).

\bibitem{McKee2005}
C.~T. McKee, S.~C. Clark, J.~Y. Walz, and W.~A. Ducker, \enquote{Relationship
  between scattered intensity and separation for particles in an evanescent
  field,} Langmuir \textbf{21}, 5783--5789 (2005).

\bibitem{Brenner1961}
H.~Brenner, \enquote{The slow motion of a sphere through a viscous fluid
  towards a plane surface,} Chem. Eng. Sci. \textbf{16}, 242--251 (1961).

\bibitem{Oetama2005}
R.~J. Oetama and J.~Y. Walz, \enquote{A new approach for analyzing particle
  motion near an interface using total internal reflection microscopy,} J.
  Colloid. Interface Sci. \textbf{284}, 323--331 (2005).

\bibitem{Carbajal-Tinoco2007}
M.~D. Carbajal-Tinoco, R.~Lopez-Fernandez, and J.~L. Arauz-Lara,
  \enquote{Asymmetry in colloidal diffusion near a rigid wall,} Phys. Rev.
  Lett. \textbf{99}, 138,303 (2007).

\bibitem{Volpe2007B}
G.~Volpe, G.~Kozyreff, and D.~Petrov, \enquote{Backscattering position
  detection for photonic force microscopy,} J. Appl. Phys. \textbf{102},
  084,701 (2007).

\end{thebibliography}
\end{document}